**Archaeology in a Vacuum: Obstacles to and Solutions for Developing a Real Space Archaeology**

Alice C. Gorman and Justin St. P. Walsh

Outer space was occupied by the human imagination long before technology enabled actual visitation to these regions. For many early commentators, planets and stars were inhabited just like Earth; however, these aliens were frequently invented to make a philosophical or satirical point (for example, in Lucian of Samosata's *True History* (second century CE) or Voltaire's *Micromegas* (1752 CE)). Galileo Galilei's observations of the near planets of the solar system with his new telescope in the 1600s was the first application of remote sensing to planetary surfaces. In the late 1800s, Percival Lowell's observations of "canals" on Mars raised hopes that sentient solar system neighbours would soon be in communication with Earth -- or invading it. Until the advent of space travel in the 1950s, the need for the as-yet non-existent professions of space archaeology (the study of human material culture in space) and xenoarchaeology (the study of non-terrestrial material culture) was equally balanced. As the 1960s and 1970s passed, not only were humans sent to the Moon, but deep space probes flew by all of the planets from Mercury to Neptune. The results were disappointing. Where there were solid surfaces, there were no canals or pyramids hinting at "civilisations" beyond our world, past or contemporary. The archaeological record of the solar system was confined to human activities.

Space archaeology is defined as the study of "the material culture relevant to space exploration that is found on Earth and in outer space (*i.e.*, exoatmospheric material) and that is clearly the result of human behavior" (Darrin and O'Leary 2009:5). Space archaeology as a discipline has its roots in the 1990s, beginning with the first attempts to describe sites and cultures associated with human space exploration. Human activity in space is worthy of archaeological research because it represents our attempts to live in an environment for which we are not evolutionarily adapted. The variable success of the technological and other engineering efforts to make the space environment survivable, habitable, and even comfortable for humans, also has its own impacts on human behavior.

The general public, and occasionally archaeological peers, sometimes struggle with the idea that archaeologists can study activities that have happened within our lifetimes; the association of space travel with futurism only intensifies this perception. As the other projects described in this volume also make clear, however, the archaeology of the contemporary world is rich with possibilities. Like our colleagues in this volume, we have been inspired by work such as the Tucson Garbage Project, which examined household waste and interviewed the residents of homes whose trash was sampled. William Rathje's team showed that merely interviewing people about their discard habits did not reveal their practices the way archaeological sampling of their trash did (Rathje and Murphy 1992). For example, Latina mothers who participated in Rathje's investigation stated that they made all their babies' food from scratch, when their garbage demonstrated that they discarded just as much store-bought baby food as other households. The point of this example is not to criticize anyone for how they feed their children (or what they say to researchers), but to demonstrate that people often do not want to discuss their behaviors and motivations because they conflict with the identity they want to project, or because they do not trust the investigator. Sometimes they are simply unable to articulate their motives and reasoning. Contemporary archaeology can fill that gap, as shown by Rathje and numerous other projects (e.g., Buchli 1999, Bailey et al. 2009, Arnold et al. 2012, De León 2015, Harrison and Breithoff 2017).

That archaeologists might turn their attention to the borderlands between present and future was presaged as long ago as 1967, when James Deetz, in his slim volume *Invitation to Archaeology*, noted that starships may well become the subject of future research. At this time the first human landing missions to the Moon were deep in the test phase and 1961 had already seen Yuri Gagarin's successful orbit of Earth. Yet despite the growth in terrestrial sites related to rocket launch and satellite industry, as well as off-Earth places where human material culture had come to rest, few took up this challenge. Space made a brief appearance in the modern material culture movement, which emerged in the US in the 1980s, when David J. Meltzer analyzed the National Air and Space Museum in Washington, DC, as a cultural artifact.

A decade later, Greg Fewer (2002) proposed that a heritage listing system was needed for sites on the Moon and Mars. This suggestion came in the context of a conference session exploring intersections between archaeology and science fiction but was the first serious attempt to apply terrestrial heritage principles to the off-Earth record. The first systematic archaeological study of a space site was Beth O'Leary's Lunar Legacy Project (1999-2001). She identified 106 objects left behind at the Tranquility Base site of the 1969 Apollo 11 landing. O'Leary and her team did not investigate the sociocultural significance of that material, however. She later worked with Lisa Westwood and Wayne Donaldson (who, at the time, was State Historic Preservation Officer for California) to successfully list the Tranquility Base objects as state historic resources in California and New Mexico in 2010. Their 2016 volume *Final Mission: Preserving*

*NASA's Apollo Culture* is the definitive text covering Apollo heritage from Earth to Moon. P.J. Capelotti (2010) identified the *facies* of what he termed the "Apollo culture," and created a list of sites associated with both crewed and uncrewed human interventions across the solar system, but he did so without analyzing the material culture of any particular site in detail. As Schiffer (2013) concluded in his survey of the field, establishing the credibility of the sub-field and addressing heritage concerns was the predominant focus of the first two decades of research.

We were also part of this process. Gorman (2007, 2009a) studied the landscapes of the rocket launch sites in Australia, French Guiana and Algeria, describing how European colonial projects were deeply entangled in the growth of space industry. During a surface survey of the former NASA satellite tracking station of Orroral Valley in Australia, she identified cable ties as a common artefact type across aerospace industry which had also migrated into everyday life (2016). In space, Gorman (2009b, 2015) has focused on the heritage and archaeological potential of orbital debris and the impacts of mining in off-Earth environments. Walsh (2012) has studied the consequences of international law for the preservation of heritage located in space contexts, suggesting that a new treaty is needed to provide adequate protection. He also has identified new, ephemeral characteristics of space technology, and investigated the consequences of those characteristics for future archaeology of human space exploration (2015).

Research in space archaeology to date has mapped and investigated terrestrial infrastructure such as launch sites, tracking stations, and industrial complexes, as well as museum collections; and off-Earth material culture including satellites and orbital debris, and planetary landing sites. While these studies demonstrated that space archaeology can be used to address new questions about human interactions with space, there has never been a substantive analysis of data collected from space sites. Carrying out an archaeological study of a site in space, whether in low-Earth orbit or on the surface of another planetary body, presents clear difficulties for researchers. For one thing, archaeologists typically need to be present at their site for direct data collection, usually by means of excavation or survey. In an outer-space context, it is generally impossible to visit sites in person today, due to the high costs and other technical barriers associated with space travel. An additional obstacle to archaeological studies of space sites is that all active space agencies with human missions have explicitly barred social-science researchers from participating in space crews (a position reiterated by NASA in its most recent call for new astronaut candidates in February 2020).

With a new emphasis by space agencies and corporations on multi-year missions to Mars and beyond, we saw an opportunity to show the relevance of an archaeological perspective by investigating how a space crew uses material culture to help structure and maintain a society. Indeed, a 1972 report by the National Academy of Sciences, titled *Human Factors in Long-Duration Spaceflight,* specifically referred to a spacecraft's crew as "a microsociety in a miniworld." Yet over almost 50 years since that publication, sociocultural aspects of life in space continue to be neglected in favor of physiological and psychological research. The International Space Station Archaeological Project is explicitly designed to fill this lacuna. We are extending archaeology into a new context, asking new questions, and contributing insights that have the potential to improve future mission success, in the first systematic archaeological study of a space habitat.

Space stations have come and gone over the decades, from the launch of the USSR's Salyut 1 station in 1971, to the recent de-orbit of China's Tiangong 1 in 2018. Only the International Space Station is currently on-orbit and occupied. Planning for ISS began in 1983, the project was announced to the public by Ronald Reagan during his State of the Union speech in 1984, and its first modules entered orbit in 1998. ISS is the largest spacecraft ever built, comprising multiple modules and a habitable volume of approximately 1000 m$^3$ (often compared by NASA to a five-bedroom house), with a total footprint equal to a US football field. At least two astronauts have continuously inhabited it since 9:21 AM (UTC) on 2 November 2000 – over 7,000 consecutive days as of early 2020.[1] The ISS project has involved five space agencies (NASA, Roscosmos, ESA, JAXA, and CSA), 25 national governments, countless private contractors, and at least 239 human visitors from 19 countries. The flight crew has varied between two and six people, representing from two and six nationalities at a time, and includes both men and women (although only 39 out of 239 of visitors to ISS – just 16% – have been women). It has at least another four to ten years of planned habitation remaining, at the end of which time its likely fate is to be de-orbited, as the Russian space station *Mir* was before it. Hence there is a limited opportunity to study society and material culture on board. Questions that are being addressed through the archaeological analysis of material culture on the ISS include:

- how people adapt their behaviors and tools to the specific requirements of life in space; and in particular how microgravity affects the development of ISS's society and culture;
- how a crew composed of people from different nations, with different languages and cultures, builds cohesion, accommodates each other, and manages conflict, in a milieu that is itself multicultural (built in different design and engineering traditions, with instruments and spaces labeled in different languages);

---

[1] In contrast, the Soviet/Russian *Mir* station, launched in February 1986, was occupied for almost twelve years and seven months in total, and for almost ten of those years continuously.

- gendered use of spaces and objects within ISS;
- how the sounds, smells, views, tastes, and textures associated with life on ISS affect crewmembers, and how astronauts adapt their behavior, movements, or the station itself in order to improve those experiences;
- how spaces and time are structured to negotiate surveillance and monitoring by ground staff, and interactions with the public.

The prohibitive costs and other obstacles to fieldwork on ISS necessitated the development of a new methodologies to allow observation and data acquisition by re-imagining traditional archaeological practices. Our inability to visit the site has already been mentioned above; we therefore have to rely on proxy methods to collect data. As there is no 'up' or 'down' in microgravity, we have also had to develop new ways of ascribing locations to artifacts and people. Where traditional archaeologists working on Earth can rely on latitude-longitude, or the Universal Transverse Mercator coordinate system to geolocate objects and installations, we have had to develop a system that is oriented along axes relative to the station's flight around the Earth: forward/aft, starboard/port, and zenith/nadir. We divide each module into 27 sub-spaces (*e.g.*, "forward-starboard-zenith" or "center-port-nadir") where an item can be found. Another variable, tied/loose, indicates whether an object is floating freely or attached to a surface.

Over almost 20 years of ISS occupation, NASA has archived millions of images of everyday life and work. This is a different type of archaeological research: not waiting for the actants to leave, but recording while they are actually in the process of daily activity, albeit frozen in an image. Using these images, we proposed to catalogue associations between crewmembers, spaces within the station, and objects/tools, to discover patterns of behavior the psychologists, sociologists and engineers have overlooked **(Fig. 1)**. As with other historical and contemporary archaeology, the augmentation of material evidence with the documentary and oral records offers insights into behaviour that it would be remiss to ignore. The Tucson Garbage Project demonstrated how the differences between the archaeological data and what people were willing to relate illuminated the social meanings of objects - such as the mundane baby food containers that were thrown away. Hypotheses generated by the image analysis about the role of artifacts and spaces in the constitution of the 'microsociety in a miniworld' can be tested against astronaut accounts in much the same way, through the administration of anonymized questionnaires to flight and ground crew.

The ISS is far from the ideal of a self-sustaining habitat; it is reliant on continual supply from Earth. The objects we observe on station are curated and managed through a complex inventory management system consisting of over 130,000 items. The crew interact with a depauperate assemblage compared to Earth, determined by factors such as space qualification, flammability, size, and weight. A limited number of items from ISS do come back to Earth. Since the end of the Shuttle program in 2011, up to approximately 2000 kg of items are brought back on SpaceX's Dragon capsule. All other materials sent to the space station either remain there, in use or in storage, or they are placed into other supply craft which are designed to be destroyed through the re-entry process. Working from archaeological analogy, we interpret the return of items from ISS as a form of discard process. Certainly, watching how some items are discarded from ISS provides much food for thought for the archaeologist. Since 2018, both Russian and US crew performing spacewalks have been instructed to simply throw broken equipment, such as a high-gain antenna electronics box, away from the station so that it will eventually fall into the atmosphere and burn up **(Fig. 2)**.

To properly contextualize the use of these items in space, a crucial method is observation of items returned from ISS, the practices and policies that comprise the cargo return ("de-integration") activity, and analysis of the values and meanings associated with those items. In 2018, we observed two return flights of the Dragon capsule, laden with scientific samples, broken/used equipment, and crew personal items. Through photography, video, and interviews with participants, we documented the procedures used by NASA contractors from Leidos Corporation and Jacobs Engineering Group to maintain control and care of the objects -- some of which must remain chilled to -80C throughout their return from ISS to California, Houston, and finally to customers.

We observed the handling of returned cargo from ISS on the Dragon CRS-13 and CRS-14 capsules in January and May 2018, respectively. During the observation, we visited work locations in Houston, TX, and Long Beach, CA; photographed and video-recorded the activities in which the returned items were documented and handled; and interviewed participants at all levels and various responsibilities relating to the cargo-return process **(Figs. 3-4)**. We gained insights into the choices made about which items to leave onboard ISS, which to destroy, and which to bring back to Earth. The items designated for return typically fall into one of three categories: scientific samples; broken equipment, or equipment that needs to be studied on Earth for some reason; and crew personal effects. We also documented the protocols for packing, unpacking, and monitoring various kinds of items. These items include paper that is sent up for the printer on ISS; scientific samples, such as crew blood, frozen at -80C; and the ice cream chosen by the Cold Stowage team to send to crew -- usually Snickers ice cream bars, because their shape fits the bags best. We even got a personal introduction to the smell of ISS when one of the cargo transfer bags was opened in front of us. Crew cannot wash thoroughly during their time on board, and they exercise for 2-3 hours daily -- "locker room," or "gym bag," with a touch of "doctor's office," goes a long way towards describing the smell of a human habitat in space.

Archaeological work on the ISS requires an adaptation of traditional methods. The NASA image archive provides a proxy for an actual survey of the interior spaces, but it may not be necessary to completely abandon the idea of fieldwork inside a space habitat. Astronauts have conducted an extraordinary range of physical science experiments and research aboard the International Space Station, including biology, chemistry, physics, astronomy, medicine, and climate research. Not all of the astronauts have significant scientific backgrounds, and if they do, they are frequently required to run experiments outside their area of expertise. More than one astronaut has mentioned to us in conversation that following instructions precisely is a key part of this process -- initiative is not encouraged! Generally speaking, crew are expected to be comfortable with any kind of work, and they therefore receive on-the-job instruction to enable them to successfully perform whatever kind of experiment is required, and to use whatever equipment is available. This seems an ideal setting to develop procedures for one or more crew members to carry out an archaeological survey of the interior of ISS during a future mission. While the analysis of the NASA image archive will enable a longitudinal mapping of human-object interactions, this more fine-grained survey will document aspects of life on board that fall between the photos, or are best achieved by direct fieldwork. Among the techniques we plan to use are surface sampling in various modules for the accretion of dust, hair, skin cells, oil, dirt, food, and other materials, analogous to soil sampling through excavation; audio recording to identify levels of ambient sound and the extent to which voices and other sounds carry through the architecture of ISS; photography to establish lines of sight from various positions, using the full freedom of movement afforded by microgravity; and documentation of specific public spaces such as eating areas, and, if possible, private spaces such as crew berths.

The methods described here are not the only ones which can enable archaeological study of space habitats. Other scholars are likely to conceive of different, even more creative approaches in the future. However, these methods are an important starting point for a subdiscipline which, like its subject matter, is still in its infancy. By combining the methods here, we can conduct deep research in an archaeological context where such work has never before been attempted. We may even be able to identify problems and propose solutions that will improve mission success by promoting harmonious social and cultural interactions in future long-duration spaceflight. Finally, the development of this methodology for space archaeology may also be useful for researchers considering contemporary archaeology projects in other remote locations or hostile environments such as under the deep ocean, in polar regions, in war zones, or anywhere else.

As well as providing an opportunity to study a unique culture created by the permanent occupation of a space habitat, ISSAP invites us to turn the lens back on terrestrial archaeology. To date, archaeological investigations have of necessity taken place predominantly against the background of terrestrial gravity. Because it is both ubiquitous and unavoidable in this setting, the role of gravity in shaping human bodies, material cultures and environmental interactions has largely remained unexamined in archaeology. The ISS is the first site of human habitation where it is possible to examine adaptations to another gravitational regime through the archaeological record.

Rather than taking the ISS as a special case, it may be more productive to consider it as one example of three human gravity adaptations (Earth, Earth orbit and the Moon), at the beginning of a trajectory which will likely come to include more (for example, Mars surface, Venus high atmosphere, and the variable gravity of crewed missions in deep space). Which social and material forms persist across these different gravity regimes will surely reveal as much about the nature of human existence as the investigation of our earliest ancestors.

**Figures**.

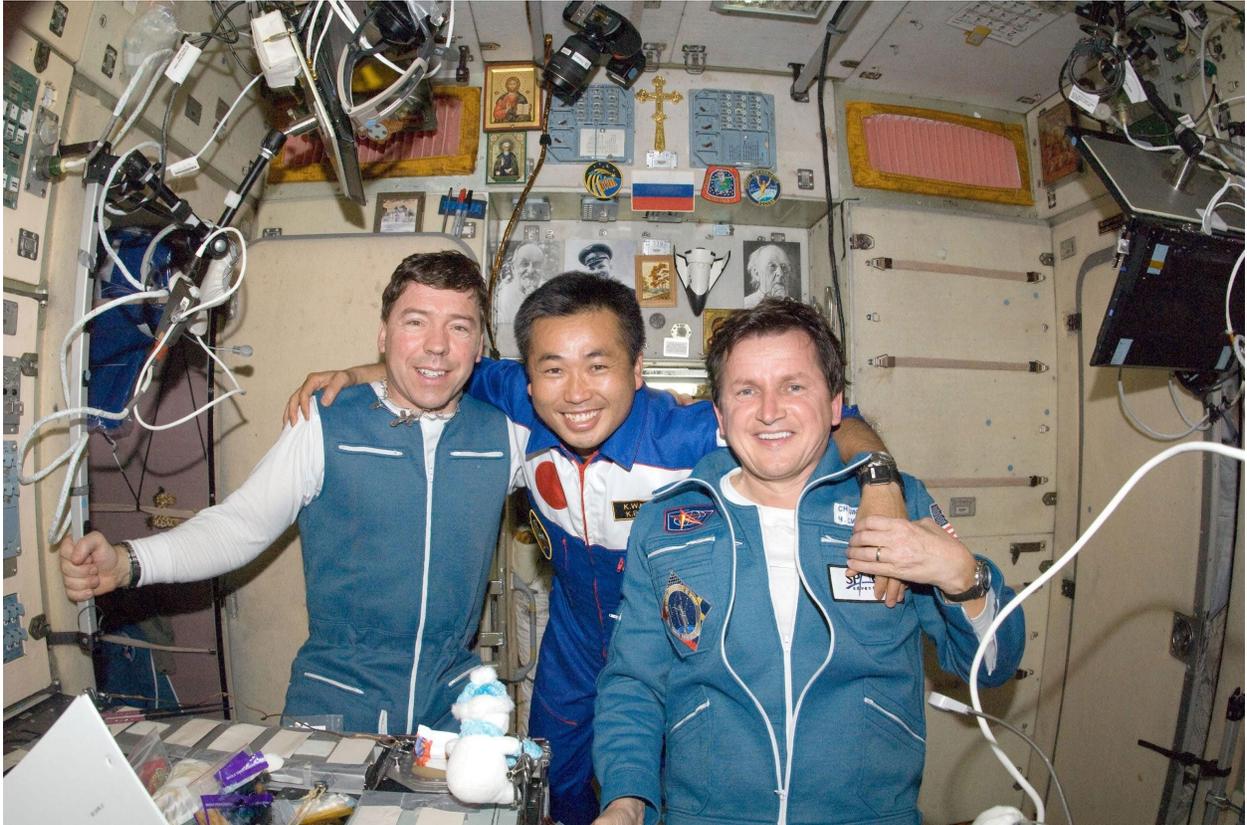

ISS018E043629

Fig. 1. ISS Expedition 19 members NASA astronaut Michael Barratt, Japan Aerospace Exploration Agency (JAXA) astronaut Koichi Wakata, and US-Hungarian spaceflight participant Charles Simonyi in the Russian Zvezda module on 28 March 2009. Also visible in the image are numerous items of material culture, including a camera, laptops, food, toys, baby wipes, four Russian Orthodox icons, a Russian Orthodox cross, a painting of the Troitse Lavra Church of St. Sergius, a Russian flag, three mission patches (Expedition 18, Soyuz TMA-13, and the Roscosmos cosmonaut corps insignia), and photographs of Soviet space heroes Konstantin Tsiolkovskiy and Yuri Gagarin. Image courtesy of NASA (posted to the Johnson Space Center Flickr page at https://www.flickr.com/photos/nasa2explore/9452806214/, accessed 28 February 2020).

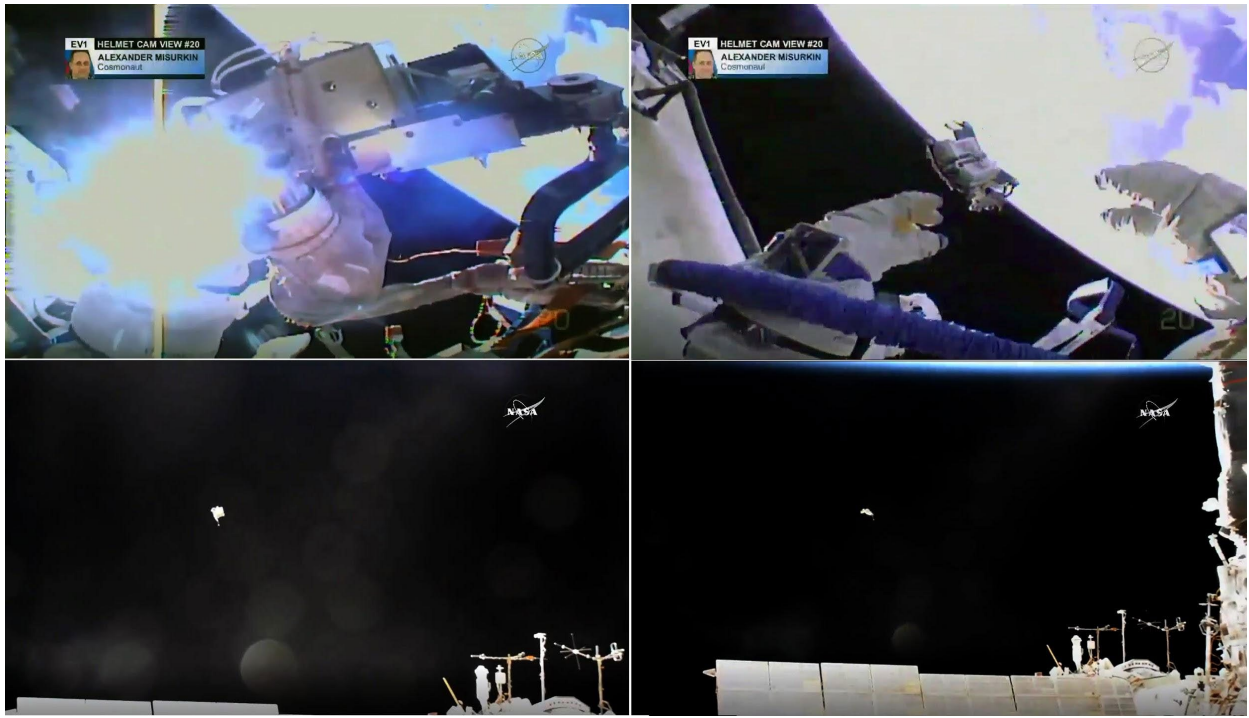

Fig. 2. Cosmonaut Alexander Misurkin discards an electronic box on 2 February 2018 by throwing it out and away from ISS. It will eventually burn up in the atmosphere. From l.-r. and top-bottom: (a) Misurkin holds the electronic box in front of him; (b) he casts it away; (c)-(d) the box is seen flying away from ISS. Screen captures from video courtesy of NASA (tweeted from the official ISS Twitter account at https://twitter.com/Space_Station/status/959514377207476224, accessed 28 February 2020).

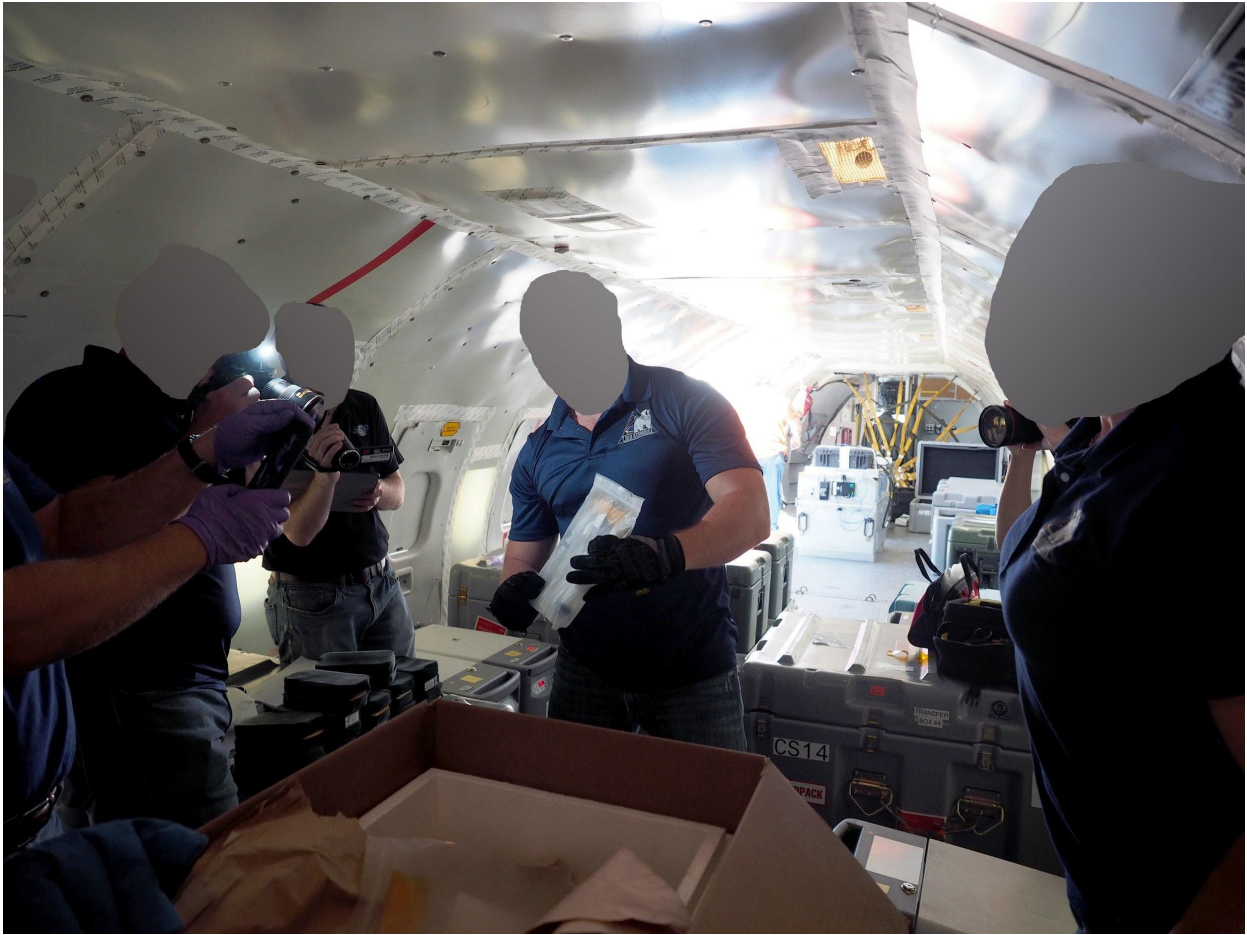

Fig. 3. Contractors from Jacobs Engineering Group document cold-stowage items returned from ISS on 7 May 2018 at Long Beach, CA, airport. Each item is photographed, video-recorded, and cross-referenced against a master list every time it is moved by the contractors or transferred to its owners. Temperatures are maintained in portable freezers down to -80C, and the temperatures are checked every 15-30 minutes while they are in the care of the Jacobs contractors. The items are contractually required to arrive in freezers at Johnson Space Center in Houston, TX, within 12 hours of being removed from SpaceX's boat at the Port of Long Beach. Photo by Justin Walsh.

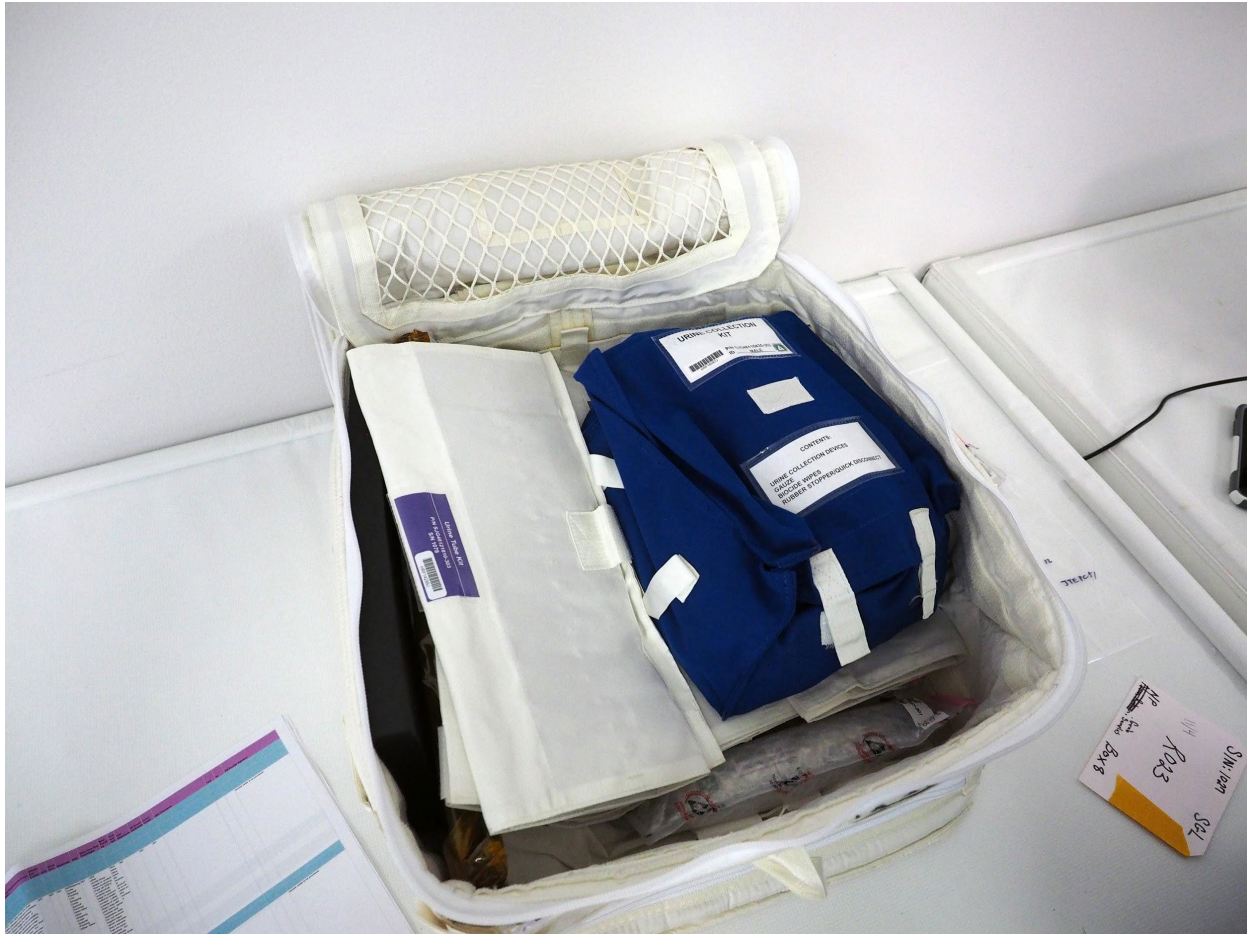

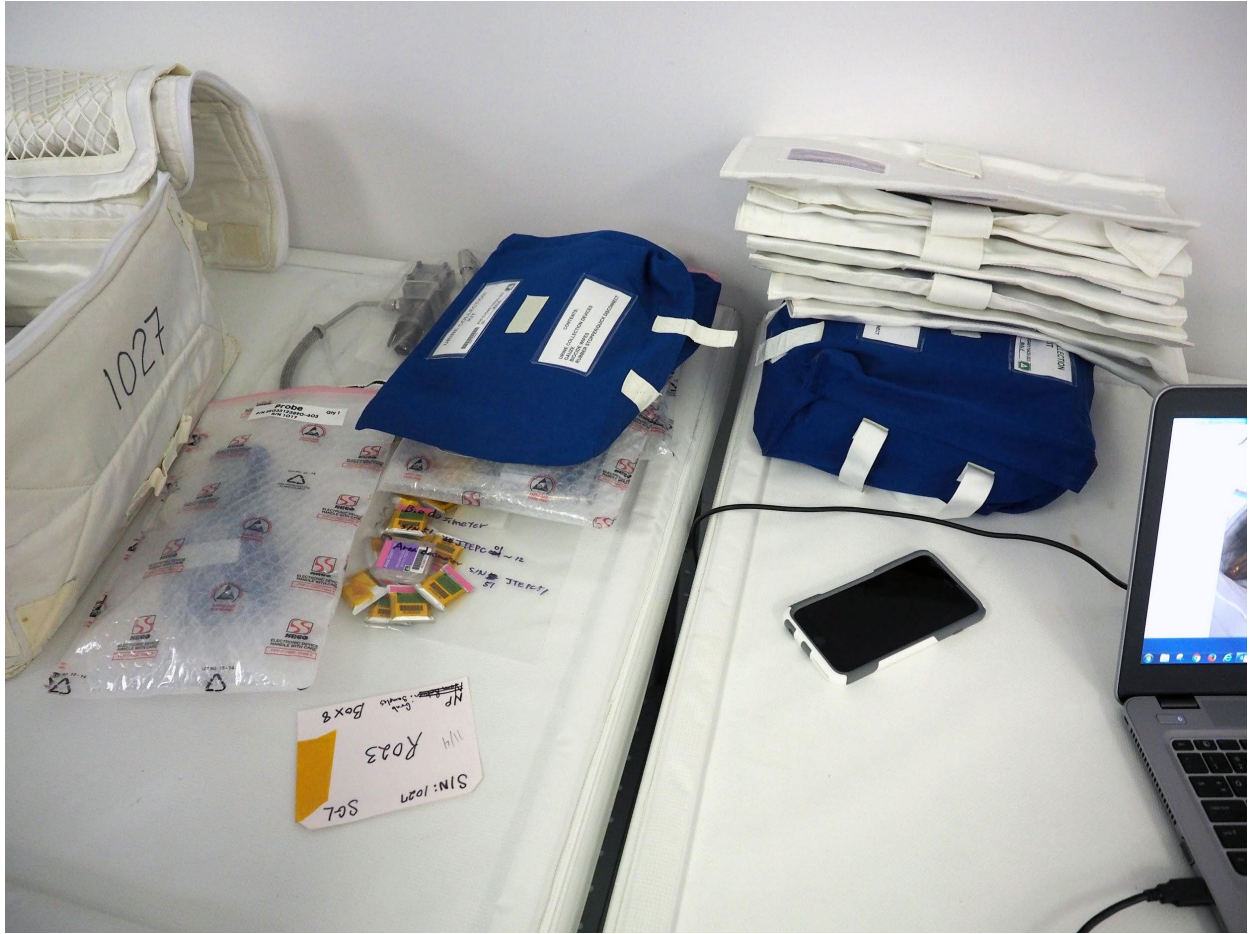

Fig. 4. (a) An ISS Cargo Transfer Bag is opened at the facilities of NASA Cargo Management Contract-holder Leidos Corporation on 19 January 2018. (b) The contents of the bag are arranged on a table waiting to be documented and recorded. Photos by Justin Walsh.